\documentclass[reprint, aps]{revtex4-2}
\usepackage[utf8]{inputenc}
\usepackage[left=2.5cm,right=2.5cm]{geometry}

\usepackage{xcolor}
\usepackage{graphicx}
\usepackage{amsmath}
\usepackage{pifont}
\usepackage{amsfonts}
\usepackage{physics}
\usepackage[mathscr]{euscript}

\newcommand{\xmark}{\ding{55}}

\begin{document}

\title{Quantum Metrology with Delegated Tasks}

\author{Nathan Shettell}
\affiliation{LIP6, CNRS, Sorbonne Université, 4 place Jussieu, 75005 Paris, France}

\author{Damian Markham}
\affiliation{LIP6, CNRS, Sorbonne Université, 4 place Jussieu, 75005 Paris, France}

\date{\today}

\begin{abstract}
    A quantum metrology scheme can be decomposed into three quantum tasks: state preparation, parameter encoding and measurements. Consequently, it is imperative to have access to the technologies which can execute the aforementioned tasks to fully implement a quantum metrology scheme. In the absence of one or more of these technologies, one can proceed by delegating the tasks to a third party. However, doing so has security ramifications: the third party can bias the result or leak information. In this article, we outline different scenarios where one or more tasks are delegated to an untrusted (and possibly malicious) third party. In each scenario, we outline cryptographic protocols which can be used to circumvent malicious activity. Further, we link the effectiveness of the quantum metrology scheme to the soundness of the cryptographic protocols.
\end{abstract}

\maketitle

\section{Motivation}

Quantum metrology has witnessed a surge in interest over the past few years \cite{giovannetti2011, degen2017}. In brief, an unknown parameter is encoded into a quantum state through some interaction; consequently, the measurement statistics of an appropriately chosen POVM will be dependent on said unknown parameter. With sufficient measurement data, an estimate of the unknown parameter can be constructed \cite{giovannetti2004, giovannetti2006, toth2014}. Quantum correlations make it possible to devise estimation strategies which attain a high level of precision, unobtainable through a classical means \cite{caves1981, bollinger1996, krischek2011, pezze2014, pezze2018}.

Fully implementing a quantum metrology scheme is technologically demanding. Quantum states must be initialized and measured with high fidelity. The quantum internet is a proposed network-like solution which can address the problem, amongst others, where parties which lack the necessary hardware can delegate the desired task to another party in the network \cite{wehner2018}. Of course, when delegating tasks, it comes with security risks; we must deal with the fact that a malicious third party could bias the estimation results or extract information for their own benefit. It is therefore imperative to take proper cryptographic precautions when delegating a portion of a metrology scheme to an untrusted third party.

\begin{figure*}[ht!]
    \centering
    \includegraphics[width=0.88\textwidth]{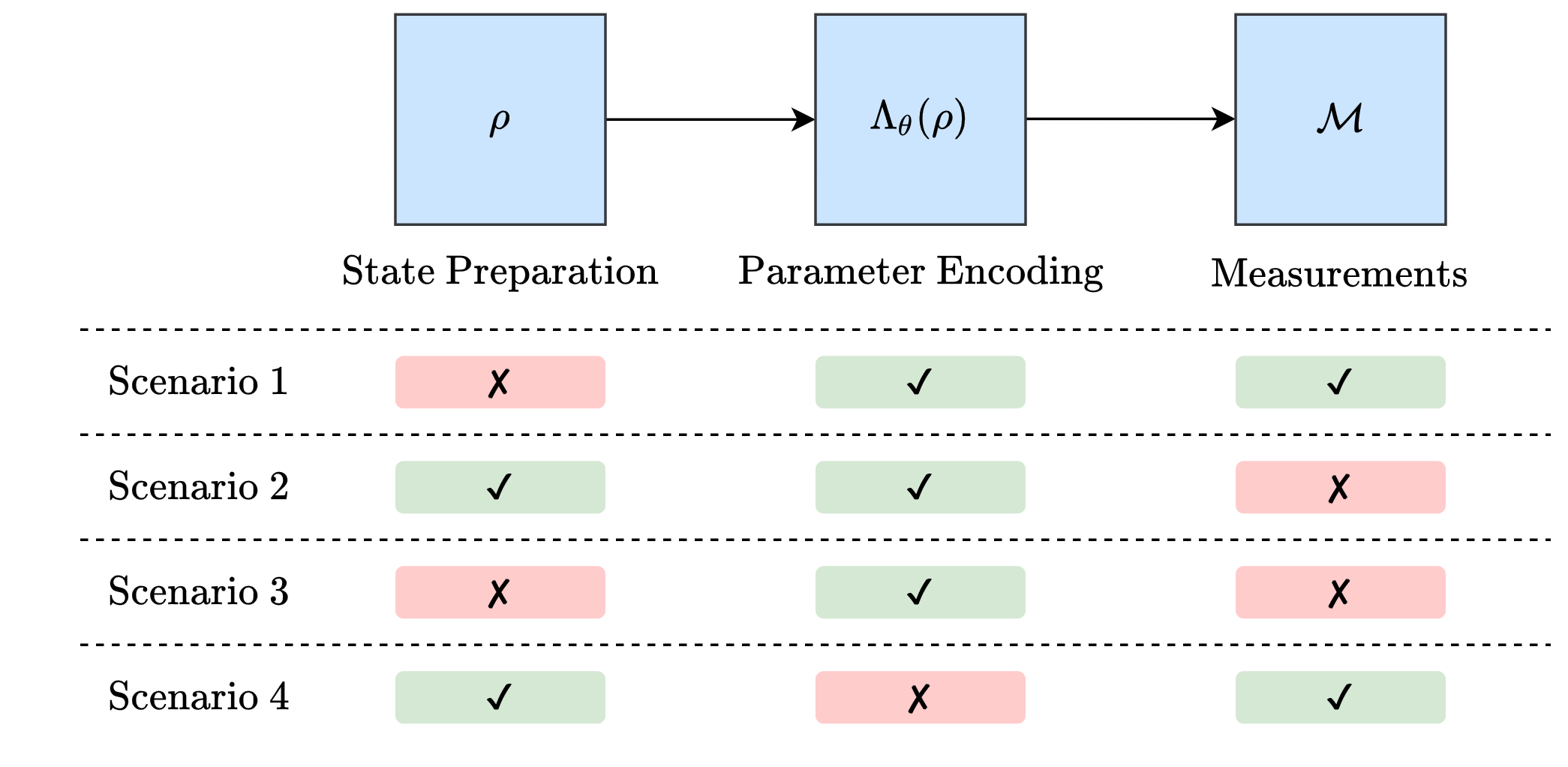}
    \caption{The different delegated quantum metrology scenarios we address in this work. A quantum metrology problem can be decomposed into three (quantum) tasks: state preparation, parameter encoding and measurements. As a whole, this is technologically demanding and it may be necessary to delegate one or more of these tasks to a third party. In this article we explore four different scenarios, each motivated through the necessity to delegate a task to a third party due to the lack of specific hardware. A red rectangle with a `\xmark' indicates that the task is delegated to a third party, as opposed to a green rectangle with a `\checkmark' which indicates that the task is not delegated. In \textit{scenario 1}, state preparation is delegated and we use verification protocols \cite{zhu2019, zhu2019general} to achieve a sense of security. In \textit{scenario 2}, the measurements are delegated and we devise an authentication based protocol to achieve a sense of security. In \textit{scenario 3}, both state preparation and measurements are delegated, and we discuss the criteria for when both of the aforementioned protocols can be used in tandem to achieve a sense of security. Finally, in \textit{scenario 4}, the parameter encoding is delegated, and we discuss the impossibility of constructing a computationally secure protocol.}
    \label{fig:DelegatedScenarios}
\end{figure*}

In the past few years, quantum cryptography has been introduced to quantum metrology to address possible security risks, such as unsecured quantum channels \cite{shettell2021, huang2019, xie2018, komar2014} and masking information from honest-but-curious eavesdroppers \cite{takeuchi2019, okane2020, yin2020}. In this work, we expand the repertoire of studied scenarios by considering the delegation of a portion of the quantum metrology process to an untrusted third party. We partition a quantum metrology problem into three tasks: state preparation, parameter encoding and measurements, and explore the repercussions when a specific task, or a combination, is delegated. The different scenarios are summarized in Fig.~(\ref{fig:DelegatedScenarios}). Note that there is an additional task of processing the measurement results and creating the estimate, however we ignore this since it is inherently a classical computation. We propose cryptographic protocols to circumvent malicious activity and achieve a sense of security for the scenarios of delegated state preparation and/or delegated measurements.

This work builds upon \cite{shettell2021}, where we introduced different quantities to measure the effectiveness of the cryptographic protocol as well as the precision of the estimate related to the quantum metrology task, namely integrity and soundness. Integrity is a measure of retaining functionality in the presence of a malicious adversary, whereas soundness provides a notion of security as it measures the ability of successfully detecting malicious activity, and thus it measures how much one can trust the resource in question. Two of the scenarios explored in this article are concerned with delegated quantum measurements, i.e. (potentially malicious) classical information, as such we have extended the mathematical definitions of integrity and soundness to allow for this possibility. Furthermore, the cryptographic protocols showcased in \cite{shettell2021} use tools from quantum message authentication schemes \cite{barnum2002, broadbent2016efficient}, in this article we show that a similar protocol can be used for the delegation of certain tasks; additionally we show that quantum state verification \cite{takeuchi2018, zhu2019, zhu2019general, markham2020} can also be adapted within cryptographic quantum metrology. Finally, in this work, we demonstrate the impossibility of delegating the task of parameter encoding in an information theoretic manner.

\section{Preliminaries}

\subsection{Soundness of a Cryptographic Protocol}

The field of quantum cryptography is extremely broad in functionality and perspectives \cite{broadbent2016, pirandola2020}. Ergo, a suitable figure of merit for a cryptographic protocol must be relevant for the scope of the protocol and provide a notion of comparability between similar protocols. In the domain of quantum verification and authentication \cite{gheorghiu2019, liu2020, broadbent2016} - for example quantum states \cite{zhu2019, zhu2019general}, quantum messages \cite{barnum2002}, or quantum computations \cite{fitzsimons2017} - a common figure of merit is soundness. The soundness of a protocol gives a notion of security as it quantifies the ability of successfully detecting alterations made by a malicious adversary and how much we can trust the resource in question. The formal mathematical definition of soundness varies depending on the formulation of the cryptographic protocol \cite{barnum2002, fitzsimons2017, zhu2019, zhu2019general, takeuchi2019Serfling}, and is sometimes referred to as verifiability \cite{gheorghiu2019}. For the sake of continuity, we use the same definition of soundness as we did in \cite{shettell2021}, which is a slightly modified version of the definition presented in \cite{barnum2002}, as they are suited to our problem, and similar statements can be made for other variants of the definition.

Verification protocols have two outputs. One is a binary accept or reject clause. The other will be a quantum state, which can be understood as either an output in its own right or an encoding of a classical measurement result (see equation (\ref{eq:measstats})). The protocols we define are equipped with ancillary qubits, which are designed to have a deterministic measurement outcome in an ideal scenario in which the untrusted party behaves as intended; if the expected measurement result is observed we assign the outcome of \textit{accept} to the protocol. However, if an unexpected result is observed, one can conclude that the untrusted third party acted maliciously and we assign the outcome of \textit{reject} to the protocol. To achieve information theoretic security, the untrusted party is assumed to be able to perform any allowable operation and is completely familiar with the protocol. In order to deal with a malicious adversary, the protocols are supported by a set of classical keys $\mathcal{K}$, where each key alters the protocol differently. A different key is chosen at random for each implementation of the protocol, and even though the adversary may have access to set of possible keys, they do not have access to the specific choice of key for any given implementation.

The formal definition of soundness is a bound on the probability of \textit{accept}, while the output quantum state $\rho_\text{out}$ is simultaneously far from the ideal output ($\rho_\text{id}$). In \cite{barnum2002} the protocol is designed $\rho_\text{id}$ being a pure state, and the distance is recorded as $\Tr( \rho_\text{id} \rho_\text{out} )$. In order to generalize this concept to mixed states, our version of soundness used the fidelity $F(\rho_\text{id}, \rho_\text{out})=(\Tr \sqrt{\sqrt{\rho_\text{id}}\rho_\text{out}\sqrt{\rho_\text{id}}})^2$. We say a protocol has soundness $\delta$ if
\begin{equation}
    \label{eq:soundness}
    \frac{1}{|\mathcal{K}|} \sum_{k \in \mathcal{K}} p_\text{acc} (k, \Gamma ) \cdot \Big( 1- F \big(\rho_\text{id}, \rho_\text{out} ( k,\Gamma ) \big) \Big) \leq \delta.
\end{equation}
Here, $\Gamma$ represents any possible attack a malicious adversary may perform, and $k \in \mathcal{K}$ is the specific key chosen. The probability of the protocol outputting \textit{accept}, $p_\text{acc}(k,\Gamma)$, and the output $\rho_\text{out}(k,\Gamma)$ are dependent on both of these quantities. Eq.~\eqref{eq:soundness} must hold for all $\Gamma$.

In the instance that $p_\text{acc} (k , \Gamma ) \geq \alpha $, then Eq.~\eqref{eq:soundness} can be written to read
\begin{equation}
    \label{eq:soundnessfidelity}
    1-\mathbb{E} \Big(  F \big(\rho_\text{id}, \rho_\text{out} \big) \Big) \leq \frac{\delta}{\alpha},
\end{equation}
where $\mathbb{E}$ denotes the expected value and we have omitted the dependence of $\rho_\text{out}$ on the key $k$ and the attack $\Gamma$ for clarity. The quantity $\alpha$ is sometimes referred to as the statistical significance \cite{zhu2019, zhu2019general}. More so, this formalization easily permits the construction of additional figures of merit which are intertwined with the soundness and statistical significance \cite{zhu2019, zhu2019general}. To connect the soundness of a cryptographic protocol to the utility of $\rho_\text{out}$ for quantum metrology, we write Eq.~\eqref{eq:soundnessfidelity} in terms of the trace distance $\mathscr{D}(\rho_\text{id}, \rho_\text{out})=\frac{1}{2} \Tr |\rho_\text{id}- \rho_\text{out}|$ \cite{shettell2021}. This is done using the arithmetic-quadratic mean inequality and the Fuchs-van de Graaf inequalities \cite{fuchs1999}
\begin{equation}
\begin{split}
    \mathbb{E} \Big(  \mathscr{D} \big(\rho_\text{id}, \rho_\text{out} \big) \Big) &\leq  \sqrt{ \mathbb{E} \Big(  \mathscr{D} \big(\rho_\text{id}, \rho_\text{out} \big)^2 \Big) } \\
    &\leq \sqrt{1-\mathbb{E} \Big(  F \big(\rho_\text{id}, \rho_\text{out} \big) \Big)} \\
    &\leq \sqrt{\frac{\delta}{\alpha}}.
\end{split}
\end{equation}

\subsection{Privacy}

Privacy is a straightforward concept which quantifies the amount of information a malicious eavesdropper can extract from a message (quantum or otherwise). The protocols outlined in this article are all completely private, which is to say that an eavesdropper can extract no information about an encoded parameter. If an eavesdropper can access the quantum state $\rho_E$, then this is achieved if
\begin{equation}
    \mathbb{E} ( \rho_E ) = \mathbb{I}/d,
\end{equation}
where $d$ is the dimension of $\rho_E$. Thus, a protocol is completely private when the expected quantum state accessible to an eavesdropper is indistinguishable from the maximally mixed state.

\subsection{Quantum Metrology}

In quantum metrology, an unknown parameter $\theta$ is encoded into an initialized quantum state $\rho$ through some CPTP map $\Lambda_\theta$; the encoded quantum state $\rho_\theta = \Lambda_\theta ( \rho)$ is then measured with respect to some POVM $\mathcal{M}$. If $\mathcal{M}$ is appropriately chosen, the measurement result will be dependent on $\theta$, and if the prepare, encode, and measure protocol is repeated sufficiently many times, $\nu \gg 1$, the measurement statistics can be used to construct an estimate $\hat{\theta}$. Formally, $\hat{\theta}$ is called an estimator and should be thought of as a function of the measurement results, the output of which is an estimate of $\theta$ \cite{kay1993}

An estimator is said to be unbiased if $\mathbb{E}(\hat{\theta})=\theta$. In classical estimation theory, the ultimate precision of an unbiased estimator is limited by the Cramér-Rao bound \cite{cramer1946}. In the realm of quantum estimation theory \cite{helstrom1969, holevo1982}, the ultimate precision is further enhanced by optimizing over all possible POVMs \cite{braunstein1994}:
\begin{equation}
    \label{eq:QCRB}
    \Delta^2 \hat{\theta} = \mathbb{E} \big( ( \hat{\theta}-\theta )^2 \big)  \geq \frac{1}{\nu \mathcal{Q}},
\end{equation}
where $\mathcal{Q}$ is the quantum Fisher information (QFI). The QFI is a measure of how much information of $\theta$ is contained within $\rho_\theta$, it is defined as
\begin{equation}
    \mathcal{Q}=\Tr \big( \rho_\theta L^2 \big),
\end{equation}
where $L$ is the symmetric logarithmic derivative which satisfies
\begin{equation}
    \partial_\theta \rho_\theta=\frac{1}{2} \big( L \rho_\theta + \rho_\theta L \big).
\end{equation}

It is always possible to saturate the quantum Cramér-Rao bound, Eq.~\eqref{eq:QCRB}, by measuring in the eigenbasis of $L$ \cite{braunstein1994}. However, this measurement is often complex and inherently dependent on $\theta$. A more practical approach is to infer $\hat{\theta}$ from an estimate of the expectation value of an observable $O$. Suppose $O$ has eigenbasis $ \{ \ket{\psi_j} \}$ with associated eigenvalues $\{ o_j \}$. If the $k$th measurement results in $\ket{\psi_j}$, by setting $m_k=o_j$ the maximum likelihood estimate \cite{toth2014} is
\begin{equation}
    \label{eq:expO}
    \expval*{\hat{O}}=\frac{1}{\nu} \sum_{k=1}^\nu m_k.
\end{equation}
The symbol $\expval*{\hat{O}}$ represents an estimate of the quantity $\expval{O}=\Tr(O \rho_\theta)$. To avoid confusion between $\expval{ \square}$ and $\mathbb{E} ( \square )$, we exclusively use $\mathbb{E} ( \square )$ for (classical) statistical quantities. The estimate of $\expval*{\hat{O}}$ can be inverted to obtain an estimate $\hat{\theta}$. By the central limit theorem, as $\nu$ increases, $\expval*{\hat{O}}$ will fluctuate closer and closer to the true value $\expval{O}$. Thus, the first order Taylor approximation
\begin{equation}
    \label{eq:TaylorApprox}
    \hat{\theta} \approx \theta + \frac{1}{|\partial_\theta \expval{O}|} \big( \expval*{\hat{O}}-\expval{O} \big)
\end{equation}
is assumed to be a valid approximation, which is used to compute the error propagation formula
\begin{equation}
    \Delta^2 \hat{\theta} = \frac{\Delta^2 \expval*{\hat{O}}}{|\partial_\theta \expval{O} |^2} = \frac{\Delta^2 O }{\nu |\partial_\theta \expval{O} |^2},
\end{equation}
where $\Delta^2 O= \Tr(O^2 \rho_\theta)-\Tr(O \rho_\theta)^2$.

Critically, quantum effects can lead to an advantage in precision compared to the best classical strategies \cite{holland1993, huelga1997}. For example by initializing an $n$ qubit GHZ state, and encoding a phase $\theta$ identically on each individual qubit, then by choosing $O=X^{\otimes n}$, one calculates $\Delta^2 \hat{\theta} = \frac{1}{\nu n^2}$ \cite{toth2014}. The quadratic scaling in $n$ is otherwise known as the Heisenberg limit and is the ultimate bound in precision allowable with quantum strategies \cite{holland1993, giovannetti2006}.

\subsection{Cryptographic Quantum Metrology}

In a cryptographic framework, many of the previously described notions from estimation theory are no longer applicable. If there is possibility that a malicious adversary tampers with any of the quantum processes (state preparation, encoding, or measurements) then there is no guarantee that the estimator will remain unbiased. Thus, there is no guarantee that the QFI is even attainable; as such the QFI is not a practical figure of merit in the realm of cryptographic quantum metrology. Instead, it is simpler to focus on a specific estimation strategy, such as the aforementioned method inferring an estimate by measuring an observable, and compare the estimate precision in the cryptographic framework to the estimate precision in the ideal framework (no malicious adversary). Because of the possibility of malicious tampering, the precision can be worse in the cryptographic setting. To fit the language of statistics, the cryptographic framework of quantum metrology injects uncertainty into the estimate. This additional uncertainty can be bounded by taking proper precautions and employing appropriate cryptographic protocols. 
For an estimate to be practical, the expected measurement statistics in the cryptographic framework must resemble the measurement statistics in the ideal framework. It will be shown that such a claim can be made by implementing appropriate cryptographic protocols. For simplicity, we restrict measurements to projection-valued measurements. We define the expected measurement statistics as a statistical ensemble $\mathcal{M}(\rho_\theta)$ (a mixed state with no coherence terms). In the ideal case, the encoded quantum state $\rho_\theta$ is measured in an orthonormal basis $\{\psi_j\}$ and the expected measurement statistics are
\begin{equation}
    \label{eq:measstats}
    \mathcal{M}(\rho_\theta) = \sum_j \dyad{\psi_j} \rho_\theta \dyad{\psi_j}.
\end{equation}
As the prepare, encode, and measure protocol is repeated $\nu$ times, the overall expected measurement statistics is $\mathcal{M}(\rho_\theta)^{\otimes \nu}$. In contrast, there is no guarantee that the expected measurement statistics are known. Further, they are not guaranteed to be dependent on $\theta$. Without loss of generality, they can be expressed as $\mathcal{M}(\rho^{\prime (k)})$ to be the statistics of the $k$th round of the prepare, encode and measure protocol. We demand that
\begin{equation}
    \label{eq:tracedistancebound}
   \frac{1}{\nu} \sum_{k=1}^\nu \mathscr{D}(\mathcal{M}(\rho_\theta), \mathcal{M}(\rho^{\prime (k)})) \leq \varepsilon,
\end{equation}
where $\mathscr{D}$ is the trace distance and $\varepsilon$ is an adjustable parameter. In \cite{shettell2021} we define a similar bound, but with respect to $\rho_\theta$ and $\rho^{\prime (k)}$; as this article explores delegated measurement, this modification is necessary. In fact, this is a stronger bound than what is presented in \cite{shettell2021} because the trace distance is contractive under CPTP maps.

For $\varepsilon \ll 1$, the most sensible strategy in the cryptographic framework is the same one as the ideal framework. That is to use the measurement results $m_1^\prime, \ldots, m_\nu^\prime$, where $\mathbb{E}(m_k^\prime) = \Tr (O \rho^{\prime (k)})=\Tr (O \mathcal{M}(\rho^{\prime (k)}))$, to construct an $\expval*{\hat{O}}^\prime$ and invert it to obtain $\hat{\theta}^\prime$. We use the notation $\square^\prime$ to indicate a quantity $\square$ in the cryptographic framework. Assuming that $\varepsilon$ is small enough such that the Taylor approximation Eq.~\eqref{eq:TaylorApprox} is still valid in the cryptographic framework, then the precision is now the sum of the variance and a bias
\begin{equation}
\begin{split}
    \Delta^2 \hat{\theta}^\prime &= \mathbb{E} \Big( \big(\hat{\theta}^\prime - \mathbb{E}(\hat{\theta}^\prime)+\mathbb{E}(\hat{\theta}^\prime)-\theta \big)^2 \Big) \\
    &= \mathbb{E} \Big( \big(\hat{\theta}^\prime - \mathbb{E}(\hat{\theta}^\prime) \big)^2 \Big)+ \big(\mathbb{E}(\hat{\theta}^\prime)-\theta \big)^2.
\end{split}
\end{equation}

As a consequence of the stronger bound Eq.~\eqref{eq:tracedistancebound} than what is presented in \cite{shettell2021}, the same proofs presented in \cite{shettell2021} hold in which we show that the bias is bounded by
\begin{equation}
    \label{eq:bias}
    \big| \mathbb{E}(\hat{\theta}^\prime)-\theta \big| \leq \frac{2 o \varepsilon}{|\partial_\theta \expval{O}|},
\end{equation}
and the integrity is bounded by
\begin{equation}
    \label{eq:integrity}
    \big| \Delta^2 \hat{\theta}^\prime-\Delta^2 \hat{\theta} \big| \leq \frac{4 o^2 (2\nu^{-1}\varepsilon+\varepsilon^2)}{|\partial_\theta \expval{O}|^2},
\end{equation}
where $o$ is the maximum magnitude of the eigenvalues of $O$. 
It follows that, in order for the metrology task to maintain a similar functionality in the cryptographic framework, the bias and variance must scale appropriately, namely,
\begin{equation}
    \mathcal{O} \big( \Delta^2 \theta^\prime \big) =\mathcal{O} \big( \Delta^2 \theta\big),
\end{equation}
for which we must have that $\varepsilon^2 \leq \nu^{-1}$.
Depending on the setting, one may relax this condition, for example, if one is primarily interested in security. In our case we will address the question of resources in the strongest case, which is to also match the scaling for accuracy.

Finally, we combine the soundness of a cryptographic protocol, Eq.~\eqref{eq:soundnessfidelity}, with the restriction on the measurement statistics Eq.~\eqref{eq:tracedistancebound}. As was previously mentioned, even if the output of a cryptographic protocol $\rho_\text{out}$ is a quantum state,  the bound on the measurement statistics is still valid because of the concavity of the trace-distance under CPTP maps. Hence, if a cryptographic protocol with soundness $\delta$ and statistical significance $\alpha$ is used in a cryptographic metrology scheme, for each prepare, encode, and measure round, then the bias, Eq.~\eqref{eq:bias}, and integrity, Eq.~\eqref{eq:integrity}, are bounded with $\varepsilon=\sqrt{\frac{\delta}{\alpha}}$, where we have made the assumption that the $\nu$ output states $\rho_\text{out}^{(1)},\ldots,\rho_\text{out}^{(\nu)}$ follow the law of large numbers:
\begin{equation}
    \frac{1}{\nu}\sum_{k=1}^\nu \mathscr{D} \big(\rho_\text{id}, \rho_\text{out}^{(k)} \big) \approx \mathbb{E} \Big(  \mathscr{D} \big(\rho_\text{id}, \rho_\text{out} \big) \Big).
\end{equation}

\section{Delegated State Preparation}

The first scenario we explore is when the task of quantum state preparation is delegated to an untrusted party. In the absence of a proper cryptographic protocol, the untrusted party could distribute any quantum state $\rho^\prime$, which could be preemptively biased to mask the true result of the parameter estimation. Fortunately, there exists a plethora of existing quantum state verification protocols \cite{takeuchi2018, pallister2018, zhu2019, zhu2019general, markham2020, liu2019, takeuchi2019Serfling}, which ensure the quantum state prepared is the desired quantum state.

Verification protocols are used to (as the name suggests) verify quantum states. Typically, this is done by requesting additional copies of the desired quantum state and by measuring the additional copies in specific bases. The measurement results are used to decide if the protocol is accepted or rejected. It should be noted that most verification protocols are tailored for specific classes of quantum states, such as graph states \cite{markham2020, takeuchi2019Serfling} or Dicke states \cite{liu2019}. More general protocols tend to require significantly more resources to achieve the same level of soundness for arbitrary quantum states \cite{takeuchi2018, pallister2018}.

As an example, consider the graph state verification protocol outlined in \cite{markham2020}. The protocol extends to any stabilizer state, which has been shown to be a useful class of states for quantum metrology \cite{shettell2020}, specifically the GHZ state which is the canonical resource for phase estimation \cite{giovannetti2006,toth2014}. The protocol takes advantage of the deterministic measurement results when measuring in a stabilizer basis \cite{fattal2004}. In summary, $N$ copies of the desired quantum state are requested, and all but one (randomly selected) is measured with respect to a random stabilizer. The protocol achieves a soundness of $\delta=1/N$. Therefore, if the verification protocol in \cite{markham2020} is incorporated into a cryptographic quantum metrology scheme, we must have that
\begin{equation}
    \frac{1}{\alpha N } \leq \frac{1}{\nu} \Rightarrow N \geq \frac{\nu}{\alpha},
\end{equation}
to maintain a similar level of precision.
After $\nu$ repetitions of the prepare, encode, and measure part of the quantum metrology scheme, this translates to a total of $\mathcal{O} \big( \nu^2 /\alpha \big)$ requested quantum states, or a quadratic increase in resources compared to the ideal framework.

\section{Delegated Measurements}

The next scenario we explore is when the measurements are delegated to an untrusted third party. A setting with an honest-but-curious adversary was explored in \cite{takeuchi2019, okane2020, yin2020} where the authors utilized tools from blind quantum computing \cite{broadbent2009} to hide the measurement results from an eavesdropper. In our version, we do not utilize the traditional blind quantum computing protocol, as it is designed solely to guarantee privacy, i.e. hide the input and output of the computation (which is the measurement in this instance) and assumes that the computation is carried out honestly. We make no assumptions about the untrusted party; for all intents and purposes the untrusted party may return arbitrary measurement results and attempt to gain information about the encoded parameter. Therefore, without proper precautions, a malicious adversary could send tailored measurement results so that the constructed estimate is a specific value of their own interest. To combat this we take inspiration from verified blind quantum computing \cite{morimae2014, fitzsimons2017} and modify the protocols we developed for performing quantum metrology over an unsecured quantum channel \cite{shettell2021} to accommodate the output being a set of measurement results.

We designate Alice as the trusted party who lacks the necessary quantum technologies to execute a quantum measurement. 
There could be several practical reasons for this. Depending on the physical systems used measurement devices themselves can be bulky, expensive affairs, such as detectors requiring cryogenic cooling, and for example Alice may be constrained to small devices, for example using optical chips so that they are portable. Furthermore, ultimately we imagine such delegation to be used in different settings in conjunction with other constraints and tasks, and so for flexibility it is prudent to consider all cases. 

\begin{figure}[!ht]
    \centering
    \includegraphics[width=0.9\linewidth]{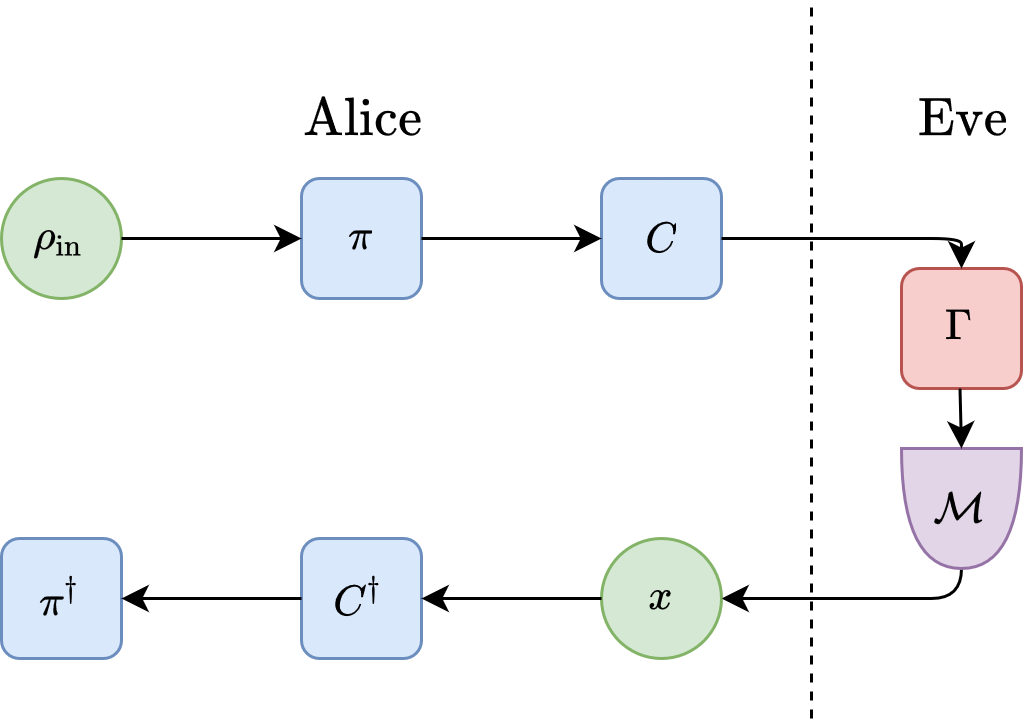}
    \caption{Before sending a quantum state to Eve to be measured, Alice can attain a sense of security by employing our protocol. In summary, Alice prepares a quantum state, $\rho_\text{in}$, which is a combination of the qubits intended from quantum metrology as well as ancillary flag qubits. Alice then encrypts the quantum state by performing a permutation $\pi$ and a Clifford operation $C$. The measurement result $x$ returned by Eve is, without loss of generality, completely arbitrary. But for all intents and purposes we write that it stems from the measurement statistics as if Eve performed the requested measurement $\mathcal{M}$ after performing an arbitrary attack $\Gamma$. Alice will perform post-processing on $x$ to correctly interpret the result, i.e., decrypt the result by inverting the Clifford operation and undoing the permutation. Alice accepts the result if the measurement result of the ancillary flag qubits corresponds to the expected result.}
    \label{fig:protocol}
\end{figure}

Alice delegates the measurement task to Eve, who will return the measurement results to Alice. Alice can then use the measurement results to construct an estimate of the unknown parameter. In an ideal setting where Eve acts honestly, Alice sends many copies of an $n$ qubit encoded quantum state $\rho_\theta$ to Eve, and requests that Eve performs a specific projective-valued measurement on each copy of of the quantum state. Eve returns the measurement results to Alice, which stems from the statistical ensemble $\mathcal{M}_\text{id}(\rho_\theta)$. In the (potentially) malicious setting, the measurement results stem from an arbitrary $\mathcal{M}_\text{id}(\rho^\prime)$. To ensure a sense of security and privacy, Alice uses a cryptographic protocol, which is described below and illustrated in Fig.~(\ref{fig:protocol}).

The protocol described below is designed solely for the case when the ideal measurement $\mathcal{M}_\text{id}$ corresponds to measuring each qubit with respect to a Pauli basis. It can be adapted to other non-entangled measurements by appropriately rotating the encryption operations. Entangled measurements could also be considered, but would require encoding over more systems. We focus on simple measurement strategies as they are the simplest to implement and the encryption strategy requires only local Clifford operations. The Clifford group is a set of unitary operations which normalize the Pauli group up to a phase of $\pm 1$. Thus, for any Clifford $C$ and $P \in \{X,Y,Z \}$
\begin{equation}
    C P C^\dagger \in \{\pm X, \pm Y, \pm Z \}.
\end{equation}
The set of locally acting Clifford operations, $\mathcal{C}_1$, can be simulated efficiently on a classical computer \cite{gottesman1998} and implemented using only sequences of $\pi/4$ rotations.

\vspace{\baselineskip}

\textbf{The Protocol:}

\begin{enumerate}
    \item Alice prepares the $m=n+t$ qubit state $\rho_\text{in}=\rho_\theta \otimes \dyad{0}^{\otimes t}$. Here, $\rho_\theta$ is the $n$ qubit quantum state where the unknown parameter $\theta$ has already been encoded, and the additional $t$ flag qubits, each initialized as $\ket{0}$, act as traps because of their deterministic measurement outcome.
    
    \item Alice encrypts $\rho_\text{in}$ by first performing a permutation $\pi$ and then applies a random Clifford $C \in \mathcal{C}_1^{\otimes m}$, $\rho_\text{in} \rightarrow \tilde{\rho}=C \pi \rho_\text{in} \pi^\dagger C^\dagger$. The permutation will insert the flag qubits at random positions so that Eve cannot distinguish between encoded qubits and flag qubits, and (as we will show) applying a random Clifford will guarantee privacy. Alice sends the permuted and encrypted quantum state to Eve.
    
    \item In the ideal case, Alice would request Eve to perform the measurement $\mathcal{M}_\text{id}$, which has Eve measuring the $n$ qubits for quantum metrology in the eigenbasis of some Pauli operator and the flag qubits in the computational basis. We write that the set of projectors which correspond to $\mathcal{M}_\text{id}$ is $\{ E \}$. In the potentially malicious case, Alice requests Eve to perform the measurement $\mathcal{M}$, which has corresponding projectors $\{C \pi E \pi^\dagger C^\dagger \}$. Doing so prevents Eve from distinguishing between a trap qubit and a qubit intended for metrology.
    
    \item Eve returns a measurement result $x$ to Alice. Without loss of generality, this measurement result originates from the measurement statistics of $\mathcal{M}(\Gamma ( \tilde{\rho}))$, where $\Gamma$ is any CPTP map which represents an attack performed by Eve.
    
    \item Alice performs classical post-processing on the measurement results to obtain the measurement results as if it had not been encrypted or permuted. When converted, the result will correspond to an outcome from the measurement statistics $\pi^\dagger C^\dagger \mathcal{M}(\Gamma(\tilde{\rho}))C \pi.$
    
    \item Alice accepts the measurement results if, after post-processing, the measurement results of the $t$ flag qubits coincided with the expected result of $\dyad{0}^{\otimes t}$. Otherwise, Alice rejects the measurement results as Eve must acted maliciously.
\end{enumerate}

The reason the protocol is designed for Pauli measurements (in the ideal case) is because a random local Clifford will map each qubit to be measured in an equal distribution of measuring in the eigenbasis of $X$, $Y$, or $Z$, as well as possibly flip the expected results. This encoding prevents Eve from distinguishing the flag qubits and the metrology qubits. As a result, the protocol is completely private, thus Eve cannot learn any information from the measurement results. The expected quantum state Eve receives is equivalent to the maximally mixed state
\begin{equation}
    \mathbb{E} (\tilde{\rho} ) = \frac{\mathbb{I}}{2^m}
\end{equation}
A proof is given in Appendix~B.

For a general measurement, it is not necessarily true that a locally acting Clifford $C$ will make the requested measurement indiscernible from the measurements on the flag qubits. The protocol can be generalized for more complex measurement strategies (e.g. measuring in a basis with inherent entanglement) by designing encryption operations in tandem with appropriately chosen flag qubits such that Eve cannot extract any information about the encoding from the requested measurement.

We show in the Appendix~B that our protocol achieves a soundness of $\delta = \frac{3n}{2t}$. Therefore, to maintain a similar level of precision in the cryptographic framework, we must have that
\begin{equation}
    \frac{3n}{2 \alpha t } \leq \frac{1}{\nu} \Rightarrow t \geq \frac{3n \nu}{2\alpha}.
\end{equation}
After $\nu$ repetitions of the prepare, encode and measure part of the quantum metrology scheme, this translates to an additional $\mathcal{O} \big( 3n \nu^2 /2\alpha \big)$ number of qubits, or a quadratic increase compared to the ideal framework.

\section{Delegated State Preparation and Measurements}

The third scenario we consider is when both the quantum state preparation and the measurements are delegated to untrusted parties. This scenario is motivated by quantum sensing networks, where a central node in the network distributes the quantum states for sensing throughout a quantum network, and encoded quantum states are returned to the central node for measurement \cite{komar2014, komar2016}. If the central node is untrusted, it is necessary for the outer nodes to incorporate a cryptographic protocol.

We continue to use the same notation introduced in the last scenario, where Alice is the trusted party and Eve is the untrusted party. Although it is plausible that the party tasked with state preparation is different than the party tasked with measurement, this distinction is irrelevant in the grand scheme of the soundness proof. Further, assuming that they are the same party results in a stronger security analysis.

In this scenario, we again restrict the requested measurement to be in a Pauli basis. We impose two additional restrictions: the first is that the requested quantum state is a stabilizer state, since they can be efficiently verified using single-qubit measurements \cite{markham2020, takeuchi2019Serfling}; the second is that the encoding map is a local unitary operation, i.e., $\Lambda_\theta \rightarrow U_\theta ^{\otimes n}$. In reality, these restrictions can be loosened. i) The requested measurement can be any single-qubit measurement scheme, and to compensate the encryption must be appropriately altered. ii) The requested quantum state can be any quantum state which can be verified using a single-qubit measurement strategy; however, without establishing the quantum state the protocol is quite vague, and it may not be possible to bound the soundness. iii) The nature of $\Lambda_\theta$ should have little to no impact on the soundness, however the third assumption is necessary to obtain a bound on the soundness. To execute the protocol, it is assumed that Alice can perform local Clifford operations.

\begin{figure}
    \centering
    \includegraphics[width=0.9\linewidth]{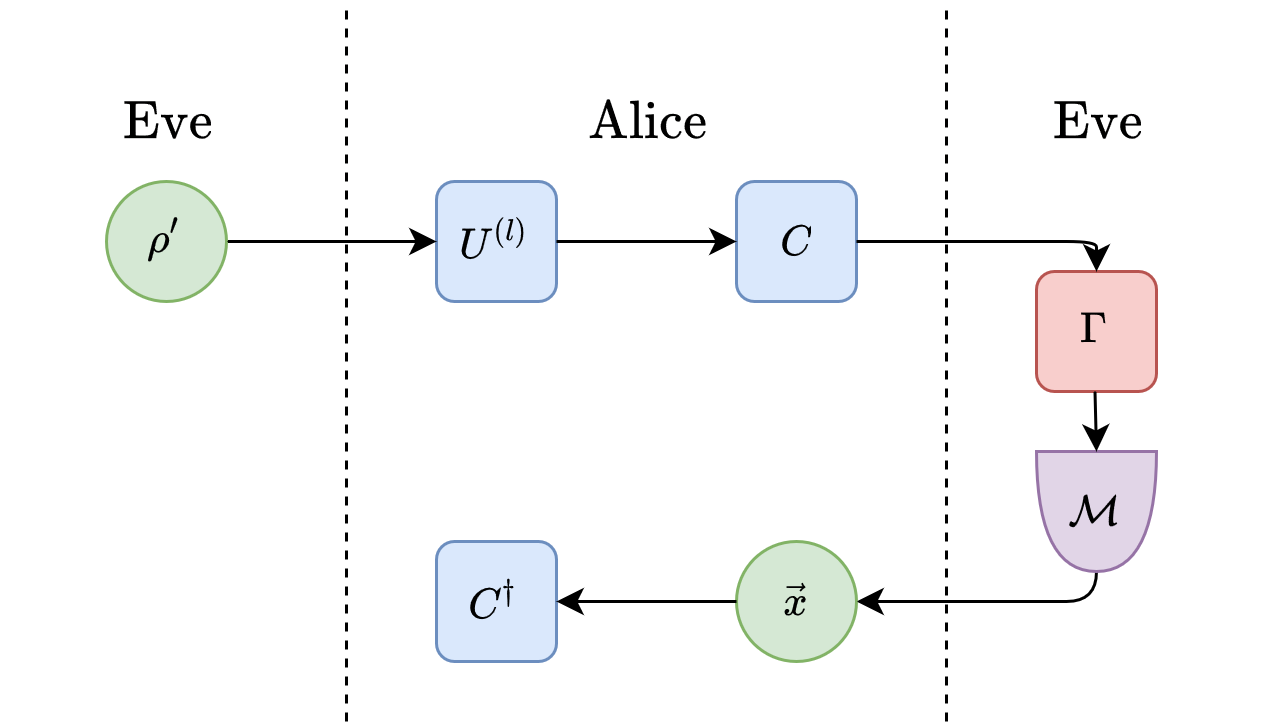}
    \caption{Ideally, Eve provides Alice the quantum state $\rho^{\otimes N}$, but in principle Eve can send Alice any $Nn$ qubit state $\rho^\prime$. Upon receipt, Alice encodes the unknown parameter of the quantum metrology problem in the $l$th block of $n$ qubits with $U^{(l)}$ and then encrypts the total quantum state with a Clifford $C$. All non-encoded blocks of qubits are then subjected to the verification protocol presented in \cite{markham2020}, where Alice requests Eve for them to be measured with respect to the basis of an arbitrary (now encrypted) stabilizer of $\rho$. Again, without loss of generality, the measurement results $\vec{x}=x_1,\ldots,x_N$ returned by Eve can be interpreted as if Eve performed an attack $\Gamma$ before performing the requested measurement $\mathcal{M}$. Alice accepts the measurement result if (after post-processing) the measurement result of the non-encoded blocks each results in a $+1$ eigenvalue with respect to their respective stabilizer measurement.}
    \label{fig:doubleprotocol}
\end{figure}

\vspace{\baselineskip}
\textbf{The Protocol:}

\begin{enumerate}
    \item Alice requests that Eve prepare $N$ copies of an $n$ qubit stabilizer state $\rho$, hence $\rho^{\otimes N}$.
    
    \item Eve sends an $Nn$ qubit state $\rho^\prime$ to Alice.
    
    \item Alice randomly chooses a positive integer $l \leq N$, this index represents the block of $n$ qubits which Alice encodes the unknown parameter onto. As the encoding map is restricted to local unitaries, this is represented by $U^{(l)}=\mathbb{I}^{\otimes n(l-1)} \otimes U_\theta^{\otimes n} \otimes \mathbb{I}^{\otimes n(N-l)}$. After encoding the unknown parameter, the quantum state Alice possesses is $U^{(l)} \rho^\prime U^{(l) \dagger}$.
    
    \item Alice randomly selects random Clifford operations $C_1, \ldots, C_N \in \mathcal{C}_1^{\otimes n}$. Alice encrypts the encoded quantum state using $C=\otimes_{j=1}^N C_j$.
    
    \item Alice randomly chooses $N-1$ stabilizers from the stabilizer group of $\rho$, $S_1,\ldots,S_{l-1},S_{l+1},\ldots,S_{N}$.
    
    \item Alice sends the encoded and encrypted quantum state to Eve for measurements. Alice requests each of the $N$ blocks of $n$ qubits to be measured with respect to a specific measurement $\mathcal{M}_j$. For $j=l$, $\mathcal{M}_l$ has corresponding projectors $\{ C_l E C^\dagger_l \}$, where $\{ E \}$ is the set of projectors of the ideal measurement. If $j \neq l$, $\mathcal{M}_j$ corresponds to measuring in the basis of $C_j S_j C_j^\dagger$ (note that if $C_j S_j C_j^\dagger$ has identity terms at certain indices then Alice requests those qubits to be measured with respect to a random Pauli basis; this will not effect the non-identity terms of the stabilizer measurement and prevent Eve from discerning between $j=l$ and $j \neq l$). For conciseness, the total measurement is labeled as $\mathcal{M}= \bigotimes_{j=1}^N \mathcal{M}_j$.
    
    \item Eve returns the measurement results $x_1,\ldots, x_N$. Without loss of generality these measurements originate from the measurement statistics of Eve performing an attack $\Gamma$ on the quantum state they receive and then performing the requested measurement $\mathcal{M}$.
    
    \item Alice performs classical post-processing to obtain the measurement results as if they had not been encrypted.
    
    \item Alice accepts the measurement results if (after post-processing) each $x_j$ with $j \neq l$ corresponds to a $+1$ eigenvalue of $S_j$. Otherwise, Alice rejects the measurement results as Eve must have acted maliciously in either the state preparation or the measurements (or both).
\end{enumerate}

In addition to the three aforementioned assumptions made with respect to this scenario, we also assume that Eve cannot alter the state between step 3 and step 4 of the protocol. This is to prevent Eve from obtaining information about $\theta$ before Alice encrypts the quantum state. With the above assumption, the reason the protocol, illustrated in Fig.~(\ref{fig:doubleprotocol}), achieves a sense of security is because in step 6, from Eve's perspective each $\mathcal{M}_j$ is indistinguishable from measuring each qubit with respect to the basis of a random Pauli. More so, even if Eve randomly guesses $l$ correctly, the measurement results are still encrypted such that Eve cannot extract any information about $\theta$. Consequently, the expected quantum state after the encryption is the maximally mixed state and thus the protocol is completely private
\begin{equation}
     \mathbb{E}\big(C U^{(l)} \rho^\prime U^{(l) \dagger} C^\dagger \big) = \frac{\mathbb{I}}{2^{Nn}},
\end{equation}
which follows from the privacy proof of the delegated measurements protocol outlined in Appendix~B.

We show in the Appendix~C that our protocol achieves a soundness of $\delta = \frac{1}{N}$. Therefore, to maintain a similar level of precision in the cryptographic framework, we must have that
\begin{equation}
    \frac{1}{\alpha N } \leq \frac{1}{\nu} \Rightarrow N \geq \frac{\nu}{\alpha}.
\end{equation}
After $\nu$ repetitions of the prepare, encode, and measure part of the quantum metrology scheme, this translates to an additional $\mathcal{O} \big( \nu^2 /\alpha \big)$ number of quantum states, or a quadratic increase compared to the ideal framework.

\section{Delegated Parameter Encoding}

The final scenario we consider is when the task of parameter estimation is delegated to an untrusted third party. From a verification perspective, the goal is to assure that some output state $\rho_\text{out}$ is close to the ideal encoded state $\rho_\theta$ with high probability. Unsurprisingly, this is an impossible task from an information theoretic standpoint without having perfect knowledge of $\theta$, which would entirely defeat the purpose of quantum metrology. The impossibility of this task stems from the fact that an adversary can manipulate the lack of information about $\theta$ to their advantage. For example, an adversary can introduce a slight bias $\Lambda_{\theta+\delta \theta}$, encode a different parameter altogether $\Lambda_{\varphi}$, encode $\theta$ into a different quantum state $\tilde{\rho}$, or do nothing at all $\mathbb{I}$. Furthermore, there is no way of guaranteeing that an adversary acts identically each round. To have security we must have some additional assumptions.

Suppose, for example,  that the abilities of the adversary are greatly limited to applying either $\Lambda_\theta$ or the identity $\mathbb{I}$. If one has a priori knowledge that $\theta \approx \theta_0$, a loose \textit{accept} criterion is for the estimate to be within some range of $\theta_0$. This `protocol' can still be manipulated by an adversary if they learn the range of acceptance: $\mathbb{I}$ is applied a small number of times such that the expected estimate falls within the acceptance range despite the added bias.

Finally, if the adversary is further hindered by assuming that they cannot access any sort of classical information - such as an a priori approximation $\theta \approx \theta_0$, or the acceptance range of the aforementioned protocol - then one can continue on with the quantum metrology scheme. This is because in this specific setting, the effective encoding map is now the CPTP map
\begin{equation}
    \rho \rightarrow (1-p)\Lambda_\theta(\rho)+p\rho,
\end{equation}
where $p$ is the effective probability that the adversary does nothing, and hence applies $\Lambda_\theta$ with effective probability $1-p$. Here, the metrology problem of estimating $\theta$ has evolved into the multiparameter problem \cite{ragy2016} of estimating $\theta$ and $p$. However, in making these assumptions, we have ventured out of the realm of cryptographic quantum metrology and into a fusion of quantum channel tomography \cite{bendersky2008} and quantum metrology.

\section{Discussion}

In this article we expanded upon the formulation of cryptographic quantum metrology \cite{shettell2021} by exploring various scenarios where a portion of a quantum metrology task is delegated to an untrusted party. In order to assure a notion of integrity, i.e., the functionality of the underlying quantum metrology problem is the same, we incorporate appropriate cryptographic protocols. For the scenarios where either state preparation or measurements are delegated to an untrusted party, we showed that cryptographic framework can attain the same level of precision as the ideal framework with a quadratic increase in resources. However, for the delegated parameter encoding scenario, we argued against the existence of any information theoretic cryptography protocols which would permit this setting. This is because any such protocol would require perfect knowledge of $\theta$, which defeats the purpose of quantum metrology.

The protocols established in this work build upon existing cryptographic protocols, namely, quantum state verification \cite{zhu2019general}, quantum message authentication \cite{barnum2002}, and blind quantum computing \cite{broadbent2009}. In principle one can incorporate other relevant cryptographic protocols, such as quantum process tomography \cite{bendersky2008, liu2020}, provided that the incorporation does not interfere with the parameter encoding. Similarly, one can incorporate protocols relevant to the specific nature of the malicious adversary; one may use a simpler protocol when dealing with honest-but-curious adversaries \cite{takeuchi2019, okane2020, yin2020}, or when dealing with specific attacks (e.g., covertness protocols, which have recently been adopted to quantum sensing \cite{bash2017, tahmasbi2021}).

For the sake of continuity with \cite{shettell2021}, we used the soundness as a cryptographic figure of merit. Note, though, that in the specific case of delegated state preparation and incorporating verification protocols, there are several possible figures of merit which are intertwined \cite{zhu2019, zhu2019general}. For example, in this article the soundness $\delta$ was bounded for a fixed $N$. However, the framework presented in \cite{zhu2019, zhu2019general} permits finding an $N$ for a fixed $\delta$ and $\alpha$. For example, for qubit stabilizer states (such as the GHZ state) the answer is $N=2(\ln2)^{-1} \delta^{-1} \ln \alpha^{-1}$ (see also \cite{unnikrishnan2020authenticated}). The bounds are different because the `worst case' attack which saturates the soundness for a fixed $N$ is different than the `worst case' attack for a fixed $\delta$.

In any of the scenarios presented, one can eliminate the possibility of a multi-round attack, i.e., a malicious attack correlated over a number of rounds, by realizing that we can equivalently formulate the problem of performing the protocol on one giant quantum state and achieve the same level of soundness with the same number of resources. For example, in the second scenario, if $\rho_\text{in} \rightarrow \rho_\text{in}^{\otimes \nu}$, then the same level of soundness is achieved since $n \rightarrow \nu n$ and $t \rightarrow \nu t$. 

In this work, as well as in \cite{shettell2021}, we restricted the quantum metrology problem to a single parameter estimation problem. However, as the cryptographic protocols do not affect the estimation strategy, one could consider multiparameter estimation problems \cite{ragy2016}. However, the estimators used in multiparameter problems are more complex and thus the bounds on the bias, Eq.~\eqref{eq:bias}, and integrity Eq.~\eqref{eq:integrity}, do not necessarily hold. Generalizing these bounds to multiparameter estimators, and even other single parameter estimators, is a future perspective for cryptographic quantum metrology.

Quantum sensing networks have recently been proposed for a variety of applications, such as synchronizing clocks \cite{komar2014, komar2016} and spatially distributed sensing problems \cite{zhuang2018, ge2018, rubio2020}. Quantum networks \cite{wehner2018, simon2017} are a collection of nodes connected via quantum channels, and different nodes have access to different quantum technologies. The work presented in this article easily integrates with quantum sensing networks to add a security aspect to the problem if one or more of the nodes are untrusted.

\textbf{Acknowledgments.} We acknowledge fruitful discussions with Elham Kashefi and financial support from the ANR through the ANR-17-CE24-0035 VanQuTe.

\bibliographystyle{unsrtnat}
\bibliography{main}

\appendix

\vfill \pagebreak \onecolumngrid

\section{Methodology on Bounding the Soundness}

\setcounter{equation}{0}
\renewcommand\theequation{A.\arabic{equation}}

In the main text, the soundness was introduced as a bound on the quantity
\begin{equation}
    \frac{1}{|\mathcal{K}|} \sum_{k \in \mathcal{K}} p_\text{acc} (k, \Gamma ) \cdot \Big( 1- F \big(\rho_\text{id}, \rho_\text{out} ( k,\Gamma ) \big) \Big),
\end{equation}
where $\rho_\text{out}(k,\Gamma)$ is the quantum state of the metrology qubits (for both protocols these are measurement statistics) conditional on the measurement results of the ancillary flag qubits resulting in \textit{accept}, and $\rho_\text{id}$ is the ideal quantum state (again measurement statistics) of the metrology qubits. This expression is introduced as it can be used to derive the integrity of the relevant quantum metrology problem, Eq.~\eqref{eq:bias} and Eq.~\eqref{eq:integrity}; however, the fidelity of quantum states, $F$, is a highly non-linear function and difficult to manipulate. Instead, we will show that the soundness can be bounded with respect to the trace of a relevant quantity (which is much simpler to manipulate).

We drop the explicit dependence on $k$ and $\Gamma$ for conciseness: $p_\text{acc}(k,\Gamma) \rightarrow p_\text{acc}$ and $\rho_\text{out} ( k,\Gamma ) \rightarrow \rho_\text{out}$. This section of the appendix is used for both protocols presented in the main text, thus the formalism is quite general; nonetheless, the specific values will be provided for clarification. We reference the first protocol as DM (delegated measurements) and the second protocol as (DSM) (delegated state preparation and measurements).

In both protocols, after post-processing the measurement result originates from the measurement statistics of
\begin{equation}
    A^\dagger \mathcal{M} \big( \Gamma( A \rho_{a} A^\dagger) \big) A,
\end{equation}
where $A$ is an encryption operation used by Alice (in DM $A \rightarrow C \pi$, in DSM $A \rightarrow C$), $\mathcal{M}$ is the measurement requested by Alice, and $\rho_a$ is the quantum state in the possession of Alice before the encryption. In both protocols, the requested measurement is some ideal projective measurement, $\mathcal{M}_\text{id}$ where the requested basis is altered with respect to the encryption $A$. Specifically, if $\mathcal{M}_\text{id}$ has projectors $\{ E \}$ then $\mathcal{M}$ has projectors $\{ A E A^\dagger \}$, thus
\begin{equation}
    A^\dagger \mathcal{M} \big( \Gamma( A \rho_{a} A^\dagger) \big) A = \mathcal{M}_\text{id} \big( A^\dagger \Gamma( A \rho_{a} A^\dagger) A \big) = \mathcal{M}_\text{id} ( \rho_f),
\end{equation}
where $\rho_f=A^\dagger \Gamma( A \rho_{a} A^\dagger) A$ is an effective final quantum state from which the measurement statistics are derived.

For the sake of clarity, we henceforth order $\mathcal{M}_\text{id}(\rho_f)$ by metrology qubits followed by the flag qubits. The measurement statistics $\mathcal{M}_\text{id}(\rho_f)$ can be expressed as a linear combination of quantum states which result in \textit{accept} or \textit{reject}
\begin{equation}
    \mathcal{M}_\text{id}(\rho_f) = p_\text{acc} \sum_{\lambda} \frac{p_\lambda}{p_\text{acc}} \rho_{\text{out},\lambda} \otimes \dyad{\lambda}+(1-p_\text{acc})\rho_{\text{disc,rej}},
\end{equation}
where $\{ \dyad{\lambda} \}$ is the set of measurement results (on the flag qubits) which are accepted by Alice, with a specific result $\dyad{\lambda}$ occurring with probability $p_\lambda$ (where $\sum_\lambda p_\lambda=p_\text{acc}$), $\rho_{\text{out},\lambda}$ is the measurement statistics of the metrology qubits if $\dyad{\lambda}$ is observed, and $\rho_{\text{disc,rej}}$ is a combination of metrology qubits and flag qubits (the form of which is irrelevant as the flag qubits result in \textit{reject} and thus the measurement statistics of metrology qubits are discarded). In DM, the only measurement result which is accepted is $\dyad{0}^{\otimes t}$; however, in DSM, the measurement result is accepted if the $j$th block of $n$ qubits is a $+1$ eigenvalue of $S_j$ for all $j \neq l$. Thus, if Alice accepts the measurement result, the measurement statistics used for quantum metrology is
\begin{equation}
    \rho_\text{out} = \sum_{\lambda} \frac{p_\lambda}{p_\text{acc}} \rho_{\text{out},\lambda}.
\end{equation}
We denote the number of accepted $\dyad{\lambda}$ as $\#\lambda$. If Alice sends Eve $\rho_\theta \otimes \dyad{\lambda}$ (for any $\lambda$), and Eve acts honestly then $\mathcal{M}_\text{id}(\rho_\theta \otimes \dyad{\lambda}) = \rho_\text{id} \otimes \dyad{\lambda}$. Using the concavity of the fidelity
\begin{equation}
\label{eq:simplify1}
\begin{split}
    \frac{1}{\# \lambda} \sum_{\lambda} F\big( \rho_\theta \otimes \dyad{\lambda},\rho_f \big) &\leq \frac{1}{\# \lambda} \sum_{\lambda} F\big( \rho_\text{id} \otimes \dyad{\lambda},\mathcal{M}_\text{id}(\rho_f) \big) \\
    &\leq F\big( \frac{1}{\# \lambda} \sum_{\lambda} \rho_\text{id} \otimes \dyad{\lambda},\mathcal{M}_\text{id}(\rho_f) \big) \\
    &= p_\text{acc} F\big( \frac{1}{\# \lambda} \sum_{\lambda} \rho_\text{id} \otimes \dyad{\lambda},\sum_{\lambda} \frac{p_\lambda}{p_\text{acc}} \rho_{\text{out},\lambda} \otimes \dyad{\lambda} \big) \\
    &= \frac{p_\text{acc}}{\# \lambda} F\big( \rho_\text{id},\sum_{\lambda} \frac{p_\lambda}{p_\text{acc}} \rho_{\text{out},\lambda} \big) \\
    &= \frac{p_\text{acc}}{\# \lambda} F\big( \rho_\text{id},\rho_{\text{out}} \big).
\end{split}
\end{equation}
Assuming that $\rho_\theta$ is a pure state, then $F\big(\rho_\theta \otimes \dyad{\lambda},\rho_f \big)=\Tr\big(\rho_\theta \otimes \dyad{\lambda}\rho_f \big)$. Because of the linearity of the trace, the summation over $\dyad{\lambda}$ can be absorbed into the trace, from which it follows that
\begin{equation}
    \Tr (\rho_\theta \otimes \Pi_\text{acc} \rho_f) \leq p_\text{acc}F\big( \rho_\text{id},\rho_{\text{out}} \big),
\end{equation}
within the DM protocol
\begin{equation}
    \Pi_\text{acc}^{(\text{DM})} = \dyad{0}^{\otimes t},
\end{equation}
and within the DSM protocol
\begin{equation}
    \Pi_\text{acc}^{(\text{DSM})} = \bigotimes_{\substack{j=0 \\ j\neq l}}^N \frac{\mathbb{I}+S_j}{2}.
\end{equation}
The probability of \textit{accept} can also be computed via
\begin{equation}
\label{eq:simplify2}
    p_\text{acc}=\Tr \big( \mathbb{I} \otimes \Pi_\text{acc} \mathcal{M}_\text{id} ( \rho_f) \big) =\Tr \big( \mathbb{I} \otimes \Pi_\text{acc} \rho_f \big).
\end{equation}
Combining Eq.~\eqref{eq:simplify1} and Eq.~\eqref{eq:simplify2}, we obtain the inequality
\begin{equation}
\label{eq:soundnessupper}
    \frac{1}{|\mathcal{K}|} \sum_{k \in \mathcal{K}} p_\text{acc} \cdot \Big( 1- F \big(\rho_\text{id}, \rho_\text{out} \big) \Big) \leq \frac{1}{|\mathcal{K}|} \sum_{k \in \mathcal{K}} \Big( \Tr \big( \mathbb{I} \otimes \Pi_\text{acc} \rho_f \big) -  \Tr \big( \rho_\theta \otimes \Pi_\text{acc} \rho_f \big) \Big) = \frac{1}{|\mathcal{K}|} \sum_{k \in \mathcal{K}}  \Tr \big( \Pi \rho_f \big),
\end{equation}
where $\Pi=(\mathbb{I}-\rho_\theta) \otimes \Pi_\text{acc}$ projects the metrology qubits of $\rho_f$ onto $\mathbb{I}-\rho_\theta$ and the flag qubits onto $\Pi_\text{acc}$. Recall that the quantum state was ordered by metrology qubits followed by flag qubits for simplicity in the derivation. The right-hand side of Eq.~\eqref{eq:soundnessupper} is much simpler to manipulate because of the linearity of the trace.

\section{Delegated Measurements to an Untrusted Party}

\setcounter{equation}{0}
\renewcommand\theequation{B.\arabic{equation}}

\textbf{Privacy:} The expected quantum state accessible to Eve is
\begin{equation}
    \mathbb{E}(\tilde{\rho}) = \frac{1}{\binom{m}{t}|\mathcal{C}_1|^m} \sum_{\pi} \sum_{C \in \mathcal{C}_1^{\otimes m}} C \pi \rho_\text{in} \pi^\dagger C^\dagger = \frac{1}{\binom{m}{t}|\mathcal{C}_1|^m 2^m} \sum_{\pi} \sum_{C \in \mathcal{C}_1^{\otimes m}} \sum_{P \in \mathcal{P}_m} \Tr ( P \pi \rho_\text{in} \pi^\dagger) CP C^\dagger,
\end{equation}
where $\mathcal{P}_m=\{ \mathbb{I},X,Y,Z \}^{\otimes m}$ is the $m$th Pauli group. Note that $\mathbb{I}$ is used to signify the identity map for any operator space, the dimension of which will be clear based on context. In the above equation, $P$ and $C$ can be constructed into $m$ local operations. Recall that for any $Q \in \{X,Y,Z \}$ the set of local Clifford operations, $\mathcal{C}_1$, will map $Q$ to a uniform distribution over $\{ \pm X, \pm Y, \pm Z \}$. Therefore, unless $P$ is uniquely equal to the identity map, the sum over $\mathcal{C}_1^{\otimes m}$ will result in zero. Thus, the summation can be simplified to
\begin{equation}
    \mathbb{E}(\tilde{\rho}) = \frac{1}{\binom{m}{t}|\mathcal{C}_1|^m 2^m} \sum_{\pi} \sum_{C \in \mathcal{C}_1^{\otimes m}} \Tr ( \pi \rho_\text{in} \pi^\dagger) C\mathbb{I}C^\dagger=\mathbb{I}/2^m.
\end{equation}

\vspace{\baselineskip}

\noindent\textbf{Local Clifford Twirling:} Before deriving a bound on the soundness of the protocol, we introduce a twirling lemma used in the protocol. The Clifford twirling lemma \cite{dankert2009} states that for any $m$ qubit quantum state $\rho$ and $Q,R \in \mathcal{P}_m$ such that $Q \neq R$, then
\begin{equation}
    \sum_{C \in \mathcal{C}_m} C Q C^\dagger \rho C R C^\dagger = 0.
\end{equation}
As our protocol uses an arbitrary local Clifford, $C \in \mathcal{C}_1^{\otimes m}$, we show that a similar result holds. To understand why, we decompose $\rho$ into a sum over the Pauli group
\begin{equation}
    \sum_{C \in \mathcal{C}_1^{\otimes m}} C Q C^\dagger \rho C R C^\dagger = \frac{1}{2^m} \sum_{P \in \mathcal{P}_m} \sum_{C \in \mathcal{C}_1^{\otimes m}} \bigotimes_{j=1}^m \big( C_j Q_j C_j^\dagger P_j C_j R_j C_j^\dagger \big),
\end{equation}
where the subscript $j$ denotes that the operator acts on the $j$th qubit. Because each $P_j$ can be expressed as a linear combination of quantum states, a corollary of the Pauli twirling lemma is that the above sum is zero if there exists a $j$ such that $Q_j \neq R_j$. Hence if $Q \neq R$
\begin{equation}
    \label{eq:localCliffordTwirl}
    \sum_{C \in \mathcal{C}_1^{\otimes m}} C Q C^\dagger \rho C R C^\dagger = 0
\end{equation}

\vspace{\baselineskip}

\noindent\textbf{Soundness:} The soundness derivation presented here is identical to the one we use present in \cite{shettell2021}. The derivation begins by representing the attack $\Gamma$ using a Kraus decomposition
\begin{equation}
    \Gamma(\sigma) = \sum_{\alpha} A_\alpha \sigma A_\alpha^\dagger,
\end{equation}
which satisfies the completeness relationship $\sum_\alpha A_\alpha A_\alpha^\dagger=\mathbb{I}$. Each Kraus operator can be written as a sum over the Pauli operators
\begin{equation}
    A_\alpha = \sum_{Q \in \mathcal{P}_m} a_{\alpha, Q} Q,
\end{equation}
where $a_{\alpha,Q}=2^{-m}\Tr(Q A_\alpha)$. Hence
\begin{equation}
    \Gamma (\sigma) = \sum_{\alpha} \sum_{Q,R \in \mathcal{P}_m} a_{\alpha,Q} a_{\alpha,R}^* Q \sigma R,
\end{equation}
where an asterisk denotes the complex conjugate and the completeness relationship translates to
\begin{equation}
    \label{eq:completeness}
    \sum_{\alpha} \sum_{Q, \mathcal{P}_m} |a_{\alpha,Q}|^2 =1.
\end{equation}
Using this formulation, the expected final quantum state can be written as
\begin{equation}
    \frac{1}{|\mathcal{K}|}\sum_{k \mathcal{K}} \rho_f = \frac{1}{\binom{m}{t}|\mathcal{C}_1|^m} \sum_{\pi} \sum_{C \in \mathcal{C}_1^{\otimes m}} \sum_{\alpha} \sum_{Q,R \in \mathcal{P}_m} a_{\alpha,Q} a_{\alpha,R}^* \pi^\dagger C^\dagger Q C \pi \rho_\text{in} \pi^\dagger C^\dagger R C \pi,
\end{equation}
which is greatly simplified thanks to local Clifford twirling, Eq.~\eqref{eq:localCliffordTwirl}, which states that the only-non vanishing terms occur when $Q=R$
\begin{equation}
    \label{eq:tempeq1a}
    \frac{1}{|\mathcal{K}|}\sum_{k \mathcal{K}} \rho_f = \frac{1}{\binom{m}{t}|\mathcal{C}_1|^m} \sum_{\pi} \sum_{C \in \mathcal{C}_1^{\otimes m}} \sum_{\alpha} \sum_{Q \in \mathcal{P}_m} |a_{\alpha,Q}|^2 \pi^\dagger C^\dagger Q C \pi \rho_\text{in} \pi^\dagger C^\dagger Q C \pi.
\end{equation}
To more easily derive a bound on the soundness, we partition $\mathcal{P}_m$ into disjoint sets $\mathcal{P}_m^{(r)}$, with $0 \leq r \leq m$, where $r$ signifies the number of non-identity terms in a Pauli, for example $\mathbb{I} \otimes X \in \mathcal{P}_2^{(1)}$, hence
\begin{equation}
    \frac{1}{|\mathcal{K}|} \sum_{k \mathcal{K}} \rho_f = \frac{1}{\binom{m}{t}|\mathcal{C}_1|^m} \sum_{\pi} \sum_{C \in \mathcal{C}_1^{\otimes m}} \sum_{\alpha} \sum_{r=0}^m \sum_{Q \in \mathcal{P}_m^{(r)}} |a_{\alpha,Q}|^2 \pi^\dagger C^\dagger Q C \pi \rho_\text{in} \pi^\dagger C^\dagger Q C \pi.
\end{equation}
As per Eq.~\eqref{eq:soundnessupper}, the soundness can be computed by projecting the above quantum state onto
\begin{equation}
    \Pi=(\mathbb{I}-\rho_\theta) \otimes \dyad{0}^{\otimes t}.
\end{equation}
There are $\binom{m-r}{t-s}$ choices of $\pi$ such that $s \leq r$ of the non-identity terms of $\pi^\dagger C^\dagger Q  C \pi \in \mathcal{P}_m^{(r)}$ interact with $s$ of the flag qubits of $\Pi$ (and thus $r-s$ non-identity terms interact with the metrology qubits of $\Pi$). Recall that the Clifford group $C_1$ will map any $P \in \{X,Y,Z \}$ to an equal distribution over $\{ \pm X, \pm Y, \pm Z \}$. The only-non vanishing terms occur when $C^\dagger $ maps these $s$ terms exclusively onto $\pm Z$, which occurs for $3^{-s}|\mathcal{C}_1|^m$ of the local Cliffords. Finally, when $r \leq t$ and $s=r$ the trace similarly vanishes as the metrology qubits are completely unaffected. Define $s_\text{max}=r-1$ if $r \leq t$ and $s_\text{max}=t$ otherwise. Using these simplifications, we obtain
\begin{equation}
     \frac{1}{|\mathcal{K}|} \sum_{k \mathcal{K}} \Tr( \Pi \rho_f) = \sum_{\alpha} \sum_{r=1}^m \sum_{Q \in \mathcal{P}_m^{(r)}} \sum_{s=0}^{s_\text{max}} 3^{-s}|a_{\alpha,Q}|^2 \frac{\binom{m-r}{t-s}}{\binom{m}{t}} \leq \sum_{r=1}^m \sum_{s=0}^{s_\text{max}} 3^{-s} \frac{\binom{m-r}{t-s}}{\binom{m}{t}},
\end{equation}
where the inequality follows from the completeness relationship, Eq.~\eqref{eq:completeness}. Re-arranging the above sum
\begin{equation}
\begin{split}
    \frac{1}{|\mathcal{K}|}\sum_{k \in \mathcal{K}} \Tr ( \Pi\rho_f ) &\leq \frac{1}{\binom{m}{t}} \sum_{s=0}^t 3^{-s} \sum_{r=s+1}^{m} \frac{\binom{m-r}{t-s}}{\binom{m}{t}} \\
    &= \sum_{s=0}^t 3^{-s} \frac{\binom{m-s}{t-s+1}}{\binom{m}{t}} \\
    &= \frac{m-t}{t+1}\sum_{s=0}^t 3^{-s} \frac{(t+1)!(m-s)!}{(t-s+1)!m!} \\
    &= \frac{m-t}{t+1}+\frac{m-t}{t+1}\sum_{s=1}^t 3^{-s} \prod_{j=0}^{s-1} \frac{t+1-j}{m-j} \\
    &\leq \frac{m-t}{t+1}+\frac{m-t}{t+1}\sum_{s=1}^t \Big(\frac{t+1}{3m}\Big)^s \\
    &\leq \frac{3}{2} \frac{m-t}{t} \\
\end{split}
\end{equation}

\section{Delegated State Preparation and Measurements to an Untrusted Party}

\setcounter{equation}{0}
\renewcommand\theequation{C.\arabic{equation}}

\textbf{Effects of the Clifford Encoding:} Before deriving a bound on the soundness of the protocol, we first find a `closed form' expression of
\begin{equation}
    S_Q=\frac{1}{|\mathcal{C}_1|^m}\sum_{C \in \mathcal{C}_1^{\otimes m}} C Q C^\dagger \sigma C Q C^\dagger,
\end{equation}
where $\sigma$ is an $m$ qubit quantum state and $Q \in \mathcal{P}_m$.

Define $\vec{x} \in \{ 0,1 \}^{\otimes}$ to be a vector of length $m$, the entries of which correspond to the non-identity terms of $Q$. Hence, if $\vec{x}=\{x_1,\ldots,x_m \}$, then $x_j=0$ if $Q_j=\mathbb{I}$ and $x_j=1$ if $P_j \in \{ X, Y, Z \}$. The magnitude of $\vec{x}$ is $x=\sum_j x_j$. We define a partial ordering $\vec{y} \preceq \vec{x}$ which satisfies $y_j \leq x_j \; \forall j$.

In the trivial case when $x=0$, the corresponding $P$ is the identity map and thus $S_Q=\sigma$. The general form is less trivial for $x=1$, but the expression can still be simplified. Without loss of generality suppose $x_1=1$ (this is for conciseness, but it will be shown to be irrelevant). We begin by expanding $\sigma$ over the Pauli basis
\begin{equation}
\begin{split}
    S_Q &=\frac{1}{2^m|\mathcal{C}_1|^m}\sum_{C \in \mathcal{C}_1^{\otimes m}} \sum_{P \in \mathcal{P}_m} \Tr(P \sigma) \bigotimes_{j=1}^m C_j Q_j C_j^\dagger P_j C_j Q_j C_j^\dagger \\
    &= \frac{1}{3 \cdot 2^m}\sum_{R_1 \in \mathcal{P}_1 / \mathbb{I} } \sum_{P \in \mathcal{P}_m} \Tr(P \sigma) R_1 P_1 R_1 \bigotimes_{j=2}^m P_j,
\end{split}
\end{equation}
where the equality follows because the sum over the local Clifford group will map $Q_1$ onto an equal distribution of $\{ \pm X, \pm Y, \pm Z \}$. We write that $\{X,Y,Z \}= \mathcal{P}_1/\mathbb{I}$. Equivalently, the above can be written as
\begin{equation}
\begin{split}
    S_Q &= \frac{1}{3 \cdot 2^m}\Big( \sum_{R_1 \in \mathcal{P}_1} \sum_{P \in \mathcal{P}_m} \Tr(P \sigma) R_1 P_1 R_1 \bigotimes_{j=2}^m P_j - \sum_{P \in \mathcal{P}_m} \Tr(P \sigma) P_1 \bigotimes_{j=2}^m P_j \Big) \\
    &= \frac{4}{3 \cdot 2^{m-1}}  \sum_{P \in \mathcal{P}_{m-1}} \Tr(\mathbb{I} \otimes P \sigma) \frac{\mathbb{I}}{2} \otimes P - \frac{1}{3 \cdot 2^m}\sum_{P \in \mathcal{P}_m} \Tr(P \sigma) P ,
\end{split}
\end{equation}
where equality arises because $P_1 \neq \mathbb{I}$ will commute with half of $\mathcal{P}_1$ and anti-commute with the other half, thus resulting in a net sum of zero. The first sum is proportional to $\sigma$ with a partial trace over the first qubit and replaced by the maximally mixed state $\mathbb{I}/2$. We define the notation $\Tr_{\vec{x}} \sigma$ to define the quantum state where all of the qubits indexed by $\vec{x}$ are traced out and replaced by the maximally mixed state $\mathbb{I}/2$. Therefore, $\sigma=\Tr_{\vec{0}} \sigma$, where $\vec{0}$ is the zero vector. Using this notation, we obtain that for $x=1$
\begin{equation}
    \label{eq:simplification1}
    S_Q=\frac{4}{3} \Tr_{\vec{x}} \sigma - \frac{1}{3} \Tr_{\vec{0}} \sigma = \sum_{\vec{y} \preceq \vec{x}} c_{\vec{y}} \Tr_{\vec{y}} \sigma,
\end{equation}
where $c_{\vec{0}}=-1/3$ and $c_{\vec{x}}=4/3$. As $S_Q$ is a valid quantum state we have that $c_{\vec{0}}+c_{\vec{x}}=1$. Even if the form was derived for $x_1=1$, the same form would have been obtained for all $\vec{x}$ with $x=1$.

We will show using inductive reasoning that the form on the right hand side of Eq.~\eqref{eq:simplification1} will hold true for any $Q^\prime \in \mathcal{P}_m$. To do this, we first suppose that $Q^\prime=QR_j$ where the non-identity terms of $Q^\prime$ and $Q$ are indexed by $\vec{x}^\prime$ and $\vec{x}$ respectively, and $R_j \in \{ X, Y, Z \}$ acts on the $j$th qubit and $x_j=0$, thus $x^\prime = x+1$ and $x^\prime_j=1$. The inductive hypothesis is
that $S_Q$ can be expressed as
\begin{equation}
    S_Q=\sum_{\vec{y} \preceq \vec{x}} c_{\vec{y}} \Tr_{\vec{y}} \sigma,
\end{equation}
with $\sum_{\vec{y} \preceq \vec{x}} c_{\vec{y}}=1$. Because of the locality of the summation over the locally acting Clifford operators we have that
\begin{equation}
\begin{split}
    S_{Q^\prime} &= \frac{1}{|\mathcal{C}_1|^m}\sum_{C \in \mathcal{C}_1^{\otimes m}} C Q^\prime C^\dagger \sigma C Q^\prime C^\dagger \\
    &= \frac{1}{|\mathcal{C}_1|}\sum_{C \in \mathcal{C}_1} C R_j C^\dagger S_Q C R_j C^\dagger \\
    &=  \sum_{\vec{y} \preceq \vec{x}} c_{\vec{y}} \Bigg( \frac{1}{|\mathcal{C}_1|}\sum_{C \in \mathcal{C}_1} C R_j C^\dagger \Tr_{\vec{y}} \sigma C R_j C^\dagger \Bigg).
\end{split}
\end{equation}
Since each $\Tr_{\vec{y}} \sigma$ is a quantum state, and $R_j$ acts solely on the $j$th qubit, the $x=1$ results can be used. Let us denote $\vec{y}+\delta_j$ as the vector with a $1$ in the $j$th position as well as the same non-zero indices as $\vec{y}$; then
\begin{equation}
    S_{Q^\prime} =  \sum_{\vec{y} \preceq \vec{x}} c_{\vec{y}} \Bigg( \frac{4}{3} \Tr_{\vec{y}+\delta_j} \sigma - \frac{1}{3} \Tr_{\vec{y}} \Bigg) = \sum_{\vec{y}^\prime \preceq \vec{x}^\prime} c_{\vec{y}^\prime} \Tr_{\vec{y}^\prime} \sigma,
\end{equation}
where $c_{\vec{y}^\prime}=\frac{4}{3} c_{\vec{y}}$ if $\vec{y}^\prime = \vec{y}+\delta_j$ for some $\vec{y} \preceq \vec{x}$, otherwise $\vec{y}^\prime = \vec{y} \preceq \vec{x}$ and we set $c_{\vec{y}^\prime}=-\frac{1}{3} c_{\vec{y}}$. It immediately follows that $\sum_{\vec{y}^\prime \preceq \vec{x}^\prime} c_{\vec{y}^\prime}=1$. Because the desired form of $S_Q$ holds $Q$ with $x=1$, then this inductive argument will hold for any $Q$ by continuously appending another non-identity Pauli at the appropriate indices.

The reason we provide this derivation is to swap the order of the parameter encoding and the sum over local Clifford operations. In the protocol $m=Nn$ and a parameter is encoded on the $l$th block of $n$ qubits via the unitary
\begin{equation}
    U^{(l)}=\mathbb{I}^{\otimes n(l-1)} \otimes U_\theta^{\otimes n} \otimes \mathbb{I}^{\otimes n(N-l)}.
\end{equation}
It follows from the locality of $U^{(l)}$ that
\begin{equation}
\label{eq:siplificationDSM}
\begin{split}
    &\frac{1}{|\mathcal{C}_1|^m}\sum_{C \in \mathcal{C}_1^{\otimes m}} C Q C^\dagger U^{(l)} \sigma U^{(l)\dagger} C Q C^\dagger \\
    =& \sum_{\vec{y} \preceq \vec{x}} c_{\vec{y}} \Tr_{\vec{y}} (U^{(l)} \sigma U^{(l)\dagger})  \\
    =& \sum_{\vec{y} \preceq \vec{x}} c_{\vec{y}} U^{(l)} ( \Tr_{\vec{y}} \sigma ) U^{(l)\dagger} \\
    =& U^{(l)} \Bigg( \frac{1}{|\mathcal{C}_1|^m}\sum_{C \in \mathcal{C}_1^{\otimes m}} C Q C^\dagger  \sigma  C Q C^\dagger \Bigg) U^{(l)\dagger}.
\end{split}
\end{equation}

\vspace{\baselineskip}

\noindent\textbf{Soundness:} For the delegated state preparation protocol, the key $k$ is a combination of three choices: which block of $n$ qubits is encoded ($l$), the encryption operation ($C$), and the stabilizers $S_1, \ldots, S_{l-1},S_{l+1},\ldots,S_N$. For a specific key, the measurement statistics originate from
\begin{equation}
    \rho_f=C^\dagger \Gamma ( C U^{(l)} \rho^\prime U^{(l) \dagger} C^\dagger) C = \sum_{\alpha} \sum_{Q \in \mathcal{P}_{Nn}} |a_{\alpha,Q}|^2 C^\dagger Q C U^{(l)} \rho^\prime U^{(l) \dagger} C^\dagger Q C,
\end{equation}
where the CPTP map $\Gamma$ was converted to a Pauli representation and simplified using the local Clifford twirling lemma  Eq.~\eqref{eq:localCliffordTwirl}. As per Eq.~\eqref{eq:siplificationDSM}, the order of operators can be swapped
\begin{equation}
    \rho_f= \sum_{\alpha} \sum_{Q \in \mathcal{P}_{Nn}} |a_{\alpha,Q}|^2 U^{(l)} C^\dagger Q C \rho^\prime C^\dagger Q C U^{(l) \dagger}.
\end{equation}

The soundness is a bound on the quantity
\begin{equation}
    \frac{1}{|\mathcal{K}|} \sum_{k \in \mathcal{K}} \Tr ( \Pi \rho_f ) = \frac{1}{N|\mathcal{C}_1|^{Nn} |\mathcal{S}|^{N-1}} \sum_{l=1}^N \sum_{C \in \mathcal{C}_1^{\otimes Nn}} \sum_{S_1 \in \mathcal{S}} \ldots \sum_{S_{l-1} \in \mathcal{S}} \sum_{S_{l+1} \in \mathcal{S}} \ldots \sum_{S_{N} \in \mathcal{S}}  \Tr ( \Pi \rho_f ),
\end{equation}
where $\mathcal{S}$ is the set of stabilizers of $\rho$ and
\begin{equation}
    \Pi=U^{(l)}\bigg( \Big( \frac{\mathbb{I}+S_1}{2} \Big) \otimes \ldots \otimes \Big( \frac{\mathbb{I}+S_{l-1}}{2} \Big) \otimes (\mathbb{I}-\rho) \otimes \Big( \frac{\mathbb{I}+S_{l+1}}{2} \Big) \otimes \ldots \otimes \Big( \frac{\mathbb{I}+S_N}{2} \Big) \bigg) U^{(l) \dagger}
\end{equation}
One of the restrictions introduced was that $\rho$ is a stabilizer state (so that we could adopt the verification protocol constructed in \cite{markham2020}); stabilizer states exhibit many symmetries, one of which is
\begin{equation}
    \frac{1}{|\mathcal{S}|} \sum_{S \in \mathcal{S}} S = \rho,
\end{equation}
hence
\begin{equation}
\label{eq:simplification2DSM}
\begin{split}
    \frac{1}{|\mathcal{K}|} \sum_{k \in \mathcal{K}} \Tr ( \Pi \rho_f ) &= \frac{1}{N|\mathcal{C}_1|^{Nn}} \sum_{l=1}^N \sum_{C \in \mathcal{C}_1^{\otimes Nn}}  \Tr (U^{(l)} \bar{\Pi}_l U^{(l) \dagger} \rho_f ) \\
    &= \frac{1}{N|\mathcal{C}_1|^{Nn}} \sum_{l=1}^N \sum_{C \in \mathcal{C}_1^{\otimes Nn}} \sum_\alpha \sum_{Q \in \mathcal{P}_{Nn}} |a_{\alpha,Q}|^2 \Tr ( \bar{\Pi}_l  C^\dagger Q C \rho^\prime C^\dagger Q C),
\end{split}
\end{equation}
where
\begin{equation}
    \bar{\Pi}_l=\Big( \frac{\mathbb{I}+\rho}{2} \Big)^{\otimes (l-1)} \otimes (\mathbb{I}-\rho) \otimes \Big( \frac{\mathbb{I}+\rho}{2} \Big)^{\otimes (N-l)}.
\end{equation}

For any quantum state $\sigma$ and projector $P$ $\Tr(P\sigma) \leq \lambda_\text{max}(P)$, where $\lambda_\text{max}(P)$ is the largest eigenvalue of $P$. Using this fact, Eq.~\eqref{eq:simplification2DSM} can be re-arranged to obtain
\begin{equation}
\begin{split}
    \frac{1}{|\mathcal{K}|} \sum_{k \in \mathcal{K}} \Tr ( \Pi \rho_f ) &= \frac{1}{|\mathcal{C}_1|^{Nn}} \sum_{C \in \mathcal{C}_1^{\otimes Nn}} \sum_\alpha \sum_{Q \in \mathcal{P}_{Nn}} |a_{\alpha,Q}|^2 \Tr \Big( C^\dagger Q C \big( \frac{1}{N}\sum_{l=1}^N \bar{\Pi}_l \big)  C^\dagger Q C \rho^\prime \Big) \\
    & \leq \frac{1}{|\mathcal{C}_1|^{Nn}} \sum_{C \in \mathcal{C}_1^{\otimes Nn}} \sum_\alpha \sum_{Q \in \mathcal{P}_{Nn}} |a_{\alpha,Q}|^2  \lambda_{\text{max}}\Big( C^\dagger Q C \big( \frac{1}{N}\sum_{l=1}^N \bar{\Pi}_l \big)  C^\dagger Q C \Big) \\
    &= \frac{1}{|\mathcal{C}_1|^{Nn}} \sum_{C \in \mathcal{C}_1^{\otimes Nn}} \sum_\alpha \sum_{Q \in \mathcal{P}_{Nn}} |a_{\alpha,Q}|^2  \lambda_{\text{max}}\Big( \frac{1}{N} \sum_{l=1}^N \bar{\Pi}_l \Big) \\
    &= \lambda_{\text{max}}\Big( \frac{1}{N} \sum_{l=1}^N \bar{\Pi}_l \Big).
\end{split}
\end{equation}
This is much simpler to compute as each $\bar{\Pi}_l$ has the same eigenbasis: tensor products of either $\rho$ or an orthogonal (pure state) $\tilde{\rho}$ (note that there are $2^{n}-1$ different $\tilde{\rho}$, but interchanging them will not affect the eigenvalue). Consider the eigenvector of $j$ copies of $\tilde{\rho}$ and $N-j$ copies of $\rho$. The only non-vanishing terms in the sum occur when the $\mathbb{I}-\rho$ term in $\bar{\Pi}_l$ interacts with a $\tilde{\rho}$ term in the eigenvector; the eigenvalue can be computed to be $\frac{j}{N2^{j-1}}$, thus
\begin{equation}
    \frac{1}{|\mathcal{K}|} \sum_{k \in \mathcal{K}} \Tr ( \Pi \rho_f ) \leq \frac{1}{N} \cdot \max_{0 \leq j \leq N} \frac{j}{2^{j-1}} = \frac{1}{N}.
\end{equation}

\end{document}